# Frequency Measurement of an Ar[+] Laser Stabilized on Narrow Lines of Molecular Iodine at 501.7 nm

F. du Burck, C. Daussy, A. Amy-Klein, A. N. Goncharov[†],

O. Lopez, C. Chardonnet, J.-P. Wallerand*

Laboratoire de Physique des Lasers, UMR 7538 CNRS, Université Paris 13

99 av. J.-B. Clément, 93430 Villetaneuse, France

[†]permanent address: Institute of Laser Physics,

Siberian Branch of the Russian Academy of Sciences, Novosibirsk, Russia

*permanent address: BNM-INM/CNAM

292 rue Saint Martin, 75003 Paris, France

*Abstract*–A spectrometer for ultra high-resolution spectroscopy of molecular iodine at wave length 501.7 nm, near the dissociation limit is described. Line shapes about 30 kHz wide (HWHM) were obtained using saturation spectroscopy in a pumped cell. The frequency of an Ar[+] laser was locked to a hyperfine component of the R(26)62-0 transition and the first absolute frequency measurement of this line is reported.

# I. INTRODUCTION

Molecular iodine shows a rich absorption spectrum from 900 nm to 500 nm (the dissociation limit of the molecule), consisting of about 100 000 rovibrational transitions between the fundamental electronic state X ($^1\Sigma_{0^+g}$) and the excited state B ($^3\Pi_{0^+u}$). Each of these shows a hyperfine structure. The natural width of these components is determined by their radiative width and by predissociation effects. These latter result from the gyroscopic and hyperfine couplings between the excited state $^3\Pi_{0^+u}$ and the dissociative state $^1\Pi_{1u}$ [1]. Both radiative and predissociative widths decrease near the dissociation limit. In particular, direct lifetime measurements by fluorescence decay have shown that the hyperfine components of the R(26)62-0 transition, accessible with an Ar$^+$ laser at 501.7 nm, should have a natural linewidth around 10 kHz [2].

Such narrow lines are of particular interest for the realization of highly stable optical frequency standards. Moreover, a better accuracy is expected, because most of the systematic errors and some experimental errors like the effect of line distortion are proportional to the linewidth. However, in a recent study of the transitions in the region 523 nm - 498 nm, the narrowest natural linewidth was estimated to be 43 kHz [3], using experimental data obtained in a sealed cell. In a previous work, in which hyperfine components of transition R(26)62-0 were detected on a supersonic beam, we obtained a linewidth of 30 kHz [4]. We present here lineshapes with a similar width obtained in saturation spectroscopy in a pumped cell. We locked the laser frequency onto these narrow lines and carried out the first absolute frequency measurement of this transition.

# II. THE SPECTROMETER

Fig. 1 depicts the Ar$^+$/I$_2$ stabilization set-up. The single mode Ar$^+$ laser is prestabilized on a Fabry-Perot resonator mode (finesse 660, free spectral range 1 GHz) by a classic Pound-Drever technique, in order to reduce its frequency jitter. For the long-term stabilization, the prestabilized laser is locked to a hyperfine component of the R(26)62-0 transition of iodine at 501.7 nm detected with the saturation spectroscopy technique in a 50 cm long sealed cell (cell 1) (switches in position 1 in Fig. 1). The iodine pressure in the cell, of about 3.3 Pa (25 mTorr), is controlled by the thermostabilization of the cell cold finger. Probe and pump beam frequencies are shifted by 250 MHz and 80 MHz respectively by the acousto-optic modulators (AOM) AOM1 and AOM2. The error signal for the long-term stabilization of the laser is obtained by the modulation transfer technique: the pump beam is frequency modulated at 125 kHz by AOM2 and the first harmonic of the saturated absorption signal is detected with the probe beam. The correction signal for frequency stabilization is applied to a piezoceramic transducer that controls the length of the Fabry-Perot resonator.

In the long-term laser stabilization, the correction bandwidth is limited by the signal-to-noise ratio (SNR) of the detected signal. We have developed an original narrow-band controller based on an adaptive noise cancelling technique for the stabilization of the beam power in the vicinity of the modulation frequency [5]. The probe beam intensity at the input of the cell is monitored by photodiode PD3 to generate the error signal applied to the narrow-band controller. It drives the RF input power of AOM1 which is used as the actuator. The closed loop system is equivalent to a notch filter tuned to the modulation frequency. The noise rejection efficiency in a narrow-band centred at 125 kHz is illustrated in Fig. 2. It shows the spectral density of the signal of photodiode PD2 located at the output of the cell. A notch in the intensity noise near 125 kHz is clearly seen. The noise is rejected by 27 dB, only 9 dB above the electronic noise level.

Iodine spectroscopy is performed in a low-pressure cell (cell 2 in Fig. 1) in order to record narrow lines. Probe and pump beam frequencies are shifted by 250 MHz and 80 MHz respectively by AOM3 and AOM4. Then, the beams cross a telescope with a pinhole to limit the transit effect in the I$_2$ cell and to improve the wavefront quality. The beam diameter in the cell is 6 mm. The hyperfine transition is detected using saturated absorption spectroscopy using the technique of frequency modulation (FM) spectroscopy [6, 7]. The modulation frequency applied to the probe beam is 2.5 MHz, a frequency for which the technical noise of the laser is negligible. However, it is well-known that a persistent problem of FM spectroscopy is the sensitivity of the detected signal to the residual amplitude modulation (RAM) generated by the frequency or phase modulator. The detection of this RAM gives a non-zero baseline and the technical noise of the laser is transferred at the detection frequency [8]. The RAM also generates a distortion of the detected line shape [9].

We use a narrow-band controller to reject the RAM of the probe beam at 2.5 MHz [10]. The error signal is obtained from a photodiode (PD4) monitoring the probe beam intensity at the input of the cell and an AOM (AOM3) driven by the cancelling signal controls the beam intensity. In our set-up, the probe beam may be phase-modulated by an electro-optic modulator (EOM), as is usual in FM spectroscopy. But, in order to simplify the scheme, the modulation may also be applied directly to AOM3 used for the intensity stabilization as shown in Fig. 1. The beam is then frequency-modulated. In this case, although the AOM introduces a spatial modulation of the beam, a 40-50 dB rejection of the 2.5 MHz RAM component may be achieved by a careful alignment of both photodiodes PD4 and PD5 on the beams [10].

Another feature of our set-up is the low-pressure iodine cell (cell 2) for the detection of narrow lines. It is 4 m long and its diameter is 15 cm. This cell may be pumped during the experiment to minimize buffer gas and impurity effects. Indeed, the width of the studied transitions at 501.7 nm is known to be very sensitive to the impurities present in the cell [11]. An iodine crystal is formed in the cell before the experiment and the iodine pressure, in the range 0.07 Pa – 0.7 Pa (0.5 mTorr – 5 mTorr), is controlled by thermostabilization of a cold finger. This cell was used in previous work for the study of transitions between hyperfine levels of the ground rovibrational states of iodine. We obtained linewidths of 2.6 kHz in stimulated Raman spectroscopy and 2.1 kHz in resonant Rayleigh spectroscopy [12, 13]. We conclude that the upper limit of impurity pressure in the cell is less than 0.04 Pa (0.3 mTorr) and that the cell itself does not limit the resolution of our experiment.

## II. THE NARROW DETECTED LINES

One of the narrowest line detected at 501.7 nm is shown in Fig. 3. It corresponds to the hyperfine component $a_7$ of transition R(26)62-0. The probe beam was frequency-modulated at 2.5 MHz by the AOM. The pressure in cell 2 was about 0.066 Pa (0.5 mTorr). The solid curve results from the fit of experimental data with the imaginary part of a Lorentzian with a half-width at half-maximum (HWHM) of 32 kHz.

The laser frequency locked to this hyperfine component was measured using a femtosecond optical frequency comb. In order to increase the signal to noise ratio (SNR) of the error signal, we worked with a pressure of 0.33 Pa (2.5 mTorr) in the cell. For this pressure, the linewidth was HWHM = 45 kHz.

## IV. THE FREQUENCY MEASUREMENT SET-UP

For the frequency measurements at 501.7 nm, the laser frequency was locked to the hyperfine component $a_7$ of transition R(26)62-0 detected in the 4-m long low-pressure cell (cell 2) instead of the sealed cell (switches in position 2 in Fig. 1). An additional 200 Hz amplitude modulation of the pump beam was introduced and a second lock-in amplifier was used in order to eliminate the residual offset of the error signal due to the Doppler background. A small amount of $Ar^+$ laser power is directed along a polarization maintaining single mode fibre to the room where the femtosecond comb generator set-up is located.

The set-up for frequency measurements at 501.7 nm is essentially similar to the one used in a previous work for absolute frequency measurement at 514.6 nm [14]. The $Ar^+$ laser frequency locked to the hyperfine component detected in the low-pressure iodine cell is compared to a mode of a wide bandwidth optical comb generated by a Kerr-lens mode-locked Ti:Sa laser and a photonic crystal fibre. The femtosecond laser is a GigaJet laser from GigaOptics pumped by a Verdi 5 laser. The 30 nm wide emission spectrum is centred at 800 nm and the repetition rate $f_{rep}$ is about 1 GHz.

The photonic crystal fibre was provided by the Department of Physics of University of Bath [15]. Its output optical spectrum extends up to one octave from 1 μm to 500 nm.

The principle of measurement of a frequency in the visible range is explained in Fig. 4. In our set-up, the carrier envelope offset frequency $f_0$ has not to be controlled. The beat note between the visible part of the output of the photonic crystal fibre and the infrared part doubled in a LBO crystal gives a signal at frequency $f_0$. It is mixed with the beat note signal between the output of the photonic crystal fibre and the Ar$^+$ laser. The mixer output signal, at frequency $f_{Ar+} - pf_{rep}$, is applied to a frequency counter. The integer $p$ is determined unambigously by changing the repetition rate $f_{rep}$ [16]. A local oscillator is locked to a 100 MHz reference signal with a high level of stability and accuracy. This reference is received from BNM-SYRTE at Observatoire de Paris through a 43 km long optical fibre, as an amplitude modulation of a 1.5 μm optical carrier [17]. The 100 MHz modulating signal is driven by a primary standard constituted by the H-maser/Cs-fountain of Observatoire de Paris. The repetition rate of the femtosecond laser is locked to the local oscillator. This local oscillator also drives the frequency counter and all synthesizers used in the experiment.

## V. MEASUREMENT OF THE HYPERFINE COMPONENT A7 OF THE R(26)62-0 TRANSITION

Fig. 5 shows a time record of a 1 s gate counting of $f_{Ar+} - pf_{rep}$ and Fig. 6 is the relative Allan deviation calculated from these data. A deviation of $7.2 \times 10^{-13}$ is obtained at 1 s. This is limited by the Ar$^+$ laser stability, the frequency reference stability being around 10 times better [17].
This value can be compared to the stability of 5 $10^{-14}$ @1 s achieved by the best frequency standards at 532 nm using reference linewidths considerably larger [18]. We are clearly limited by the stability of our Ar$^+$ laser. The development of compact sources around 500 nm realized with frequency doubled diode-pumped solid-state lasers based on Yb$^{3+}$-doped materials [19], intrinsically more stable, should considerably improved this optical reference. An other limitation is the low signal size due to the weaker strength of the transitions and the lower pressure used to approach the natural linewitdth. The detection design might be adapted to obtained a sufficient SNR.

During a preliminary study, we estimated the frequency shifts associated to the parameters of the $Ar^+$ laser frequency stabilization set-up: the pressure in the cell seems to be the most sensitive parameter with about -38.4 kHz/Pa (-5 kHz/mTorr). This sensitivity coefficient has been measured around a pressure of 0.33 Pa (2.5 mTorr), for which the frequency shift of the transition is no more linear with the pressure and is larger than the linear frequency shift observed for higher pressure. Besides, the non linear frequency shift is proportionnal to $(\Gamma_e - \Gamma_g)/(\Gamma_e + \Gamma_g)$ where $\Gamma_e$ and $\Gamma_g$ are the excited and ground state decay rates respectively [14], and results in a larger shift for a smaller linewidth. This leads to a larger sensitivity coefficient, compared to values usually found for frequency standards based on molecular iodine. The power of the probe beam is also an influence parameter: a frequency shift of about 1 kHz for a decrease by a factor 2 of the probe beam power was observed. No shift associated to the modulation index was found.

The frequency measurement of the hyperfine component $a_7$ of the transition R(26)62-0 was repeated during a period of a few days, with about the same pressure in the cell (0.33 Pa - 2.5 mTorr), the pump beam power between 2 mW and 3.4 mW and the probe beam power between 300 µW and 500 µW. The probe beam was either frequency-modulated by the AOM or phase-modulated by the EOM. In each case, the modulation index was close to unity. Results are reported in Fig. 7. The main part of the day to day frequency reproducibility may be attributed to the pressure fluctuations in the cell due to the pumping process during the experiments. The large sensitivity to this parameter could explain the 2 kHz difference observed between 11/02/04 and 12/02/04 measurements.

The absolute frequency of the laser locked to component $a_7$ of transition R(26)62-0 in these conditions is found to be 597 366 498 654.62 kHz. The standard deviation of the mean values obtained for each of the 4 measurement sets is 0.85 kHz. On 23/01/04, five measurements were made during the day, the laser being unlocked and the optical system re-aligned between each measurement. A reproducibility of 65 Hz (~$10^{-13}$ in relative value) was then obtained, which gives an indication of the reproducibility which could be achieved with a better control of the pressure.

## VI. CONCLUSION

We have described our ultra high resolution spectrometer for molecular iodine at 501.7 nm based on an $Ar^+$ laser and a low-pressure continuously pumped cell. For hyperfine components of transition R(26)62-0, we obtained linewidths of about 30 kHz, which are, to our knowledge, the narrowest value obtained in iodine in a cell. Our $Ar^+$ laser was locked on one of these lines and the first frequency measurement of this transition was carried out.

Such transitions are clearly of particular interest for the realisation of optical frequency standards, especially with the development of compact sources around 500 nm, which will give the possibility to take advantage of the high metrological potential of the iodine transitions close to the dissociation limit.

**FIGURE CAPTIONS**

**Fig. 1:** $Ar^+$ laser stabilization scheme (FP: Fabry-Perot resonator; correct.: controller; PD: photodiode; AOM: acousto-optic modulator; switches in position 1: the laser is stabilized on cell 1 for spectroscopy; switches in position 2: the laser is stabilized on cell 2 for frequency measurements).

**Fig. 2:** Noise spectral density of the probe beam near 125 kHz for the $Ar^+$ laser long-term stabilization.

**Fig. 3:** Hyperfine component $a_7$ of R(26)62-0 transition in iodine at 501.7 nm detected in FM spectroscopy (iodine pressure: 0.066 Pa (0.5 mTorr); saturating beam power: 1.5 mW; probe beam power: 0.4 mW; beam diameter: 6 mm; time constant: 100 ms; three sweeps of 60 s).

**Fig. 4:** Principle of absolute frequency measurement in the visible range.

**Fig. 5:** Time record of 1 s gate counting of $f_{Ar+} - pf_{rep}$. (the figure shows the frequency deviation from the mean value versus time).

**Fig. 6:** Relative Allan deviation calculated from data of Fig. 5.

**Fig. 7:** Frequency measurement of hyperfine component $a_7$ of R(26)62-0. (pressure in the cell: 0.33 Pa - 2.5 mTorr; saturating beam power between 2 mW and 3.4 mW; probe beam power between 300 µW and 500 µW; on 23/01/04, the probe beam was frequency-modulated by the AOM; the other days, it was phase-modulated by the EOM; modulation index ~1).

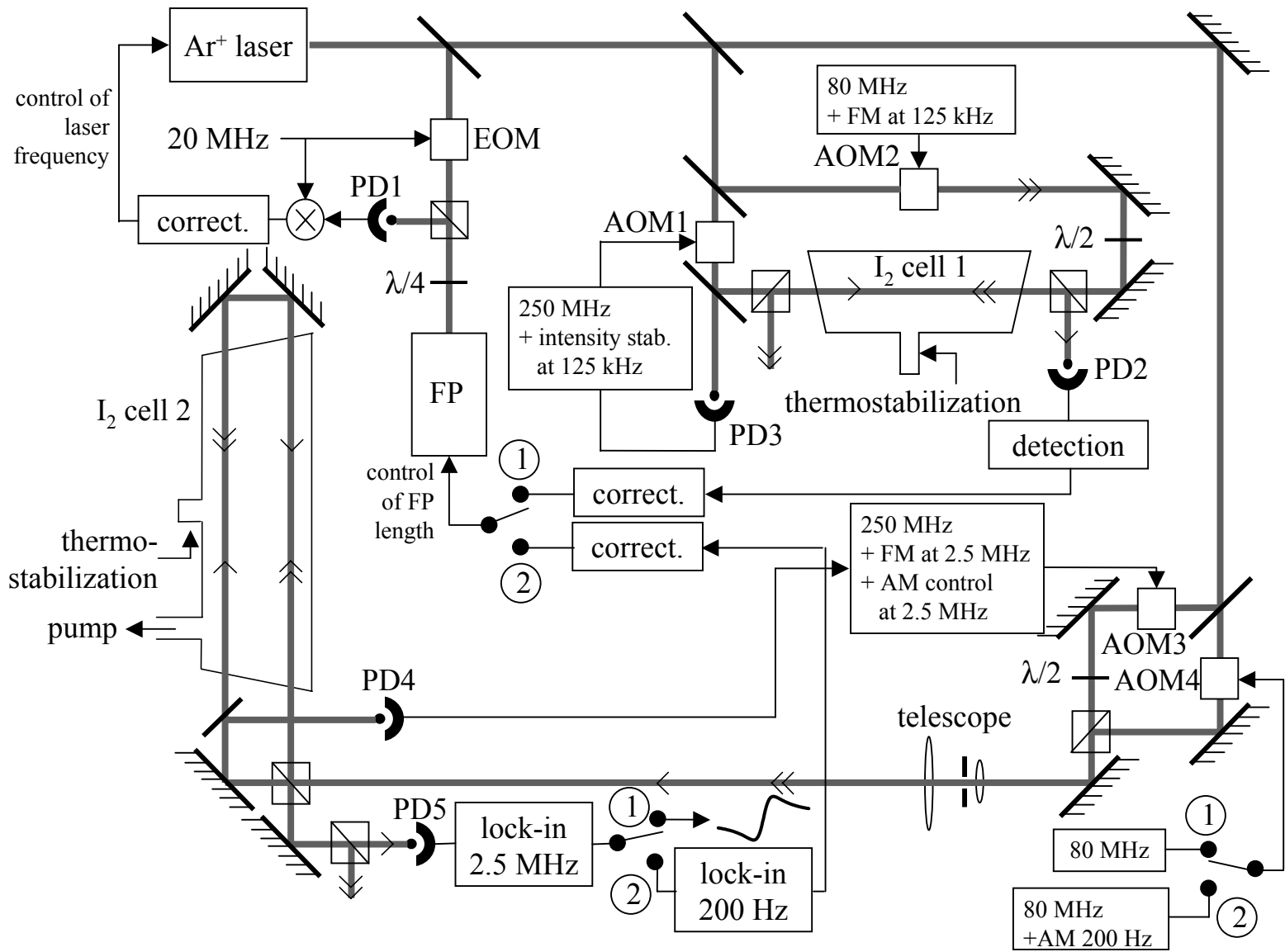

**Fig. 1**

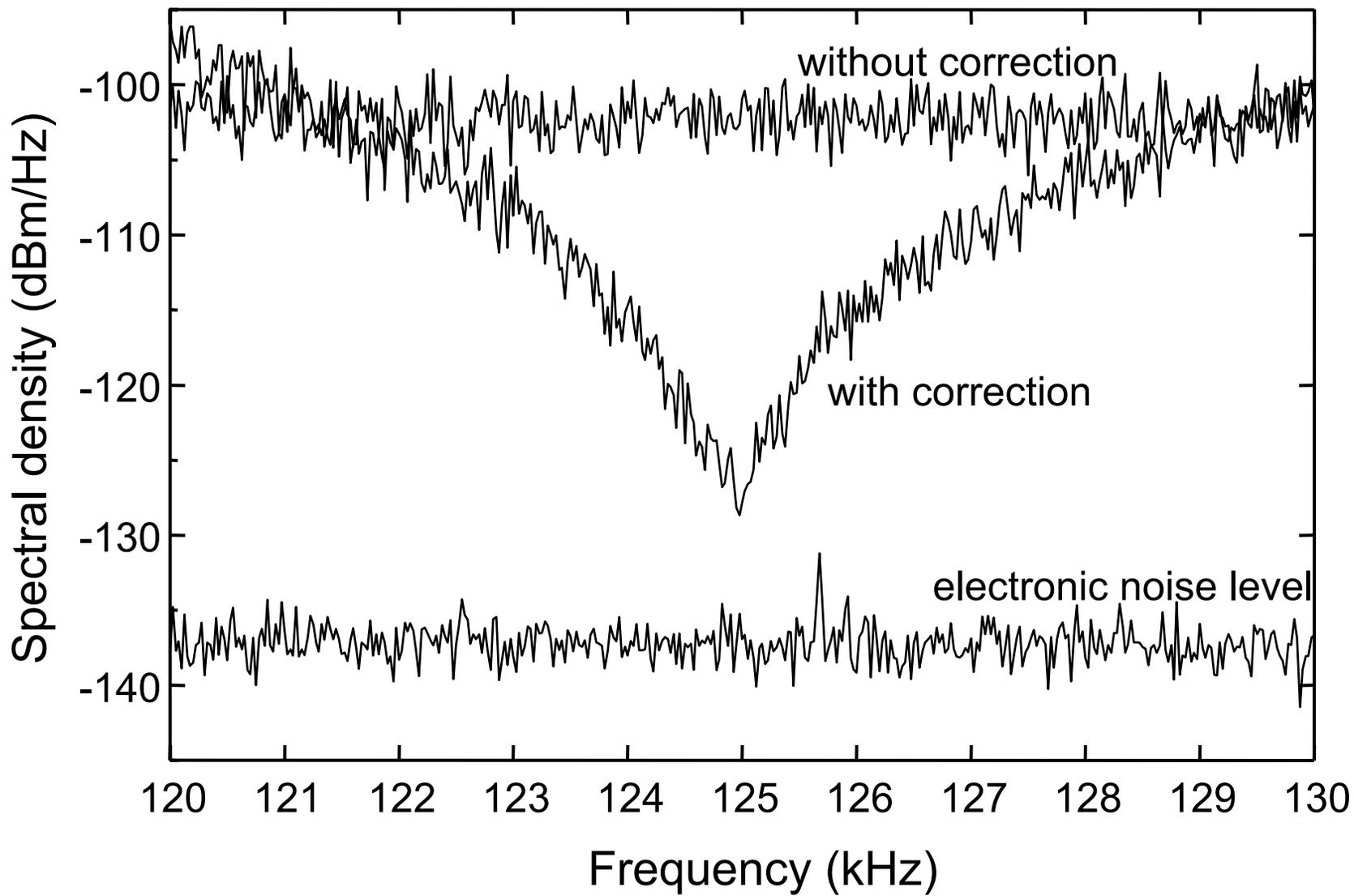

**Fig. 2**

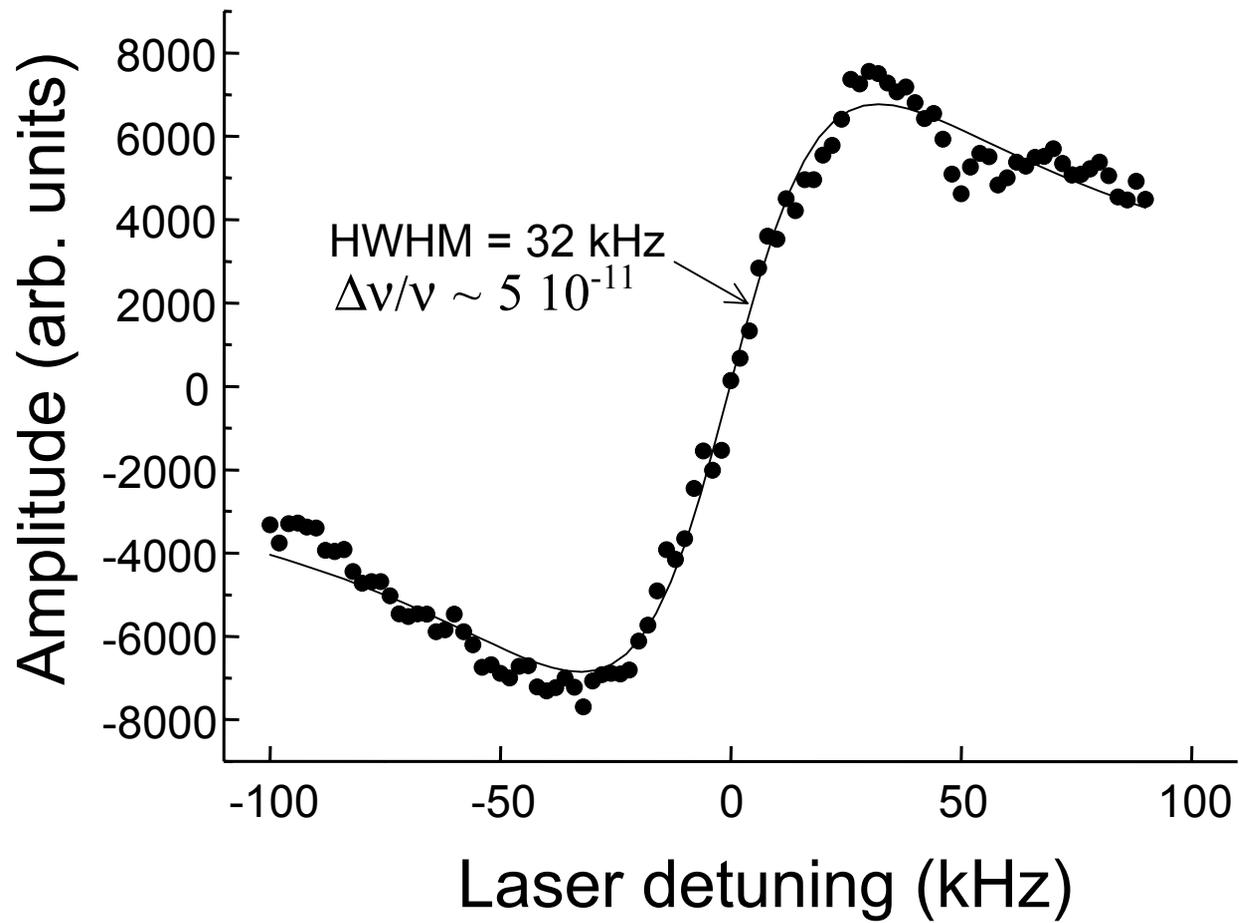

**Fig. 3**

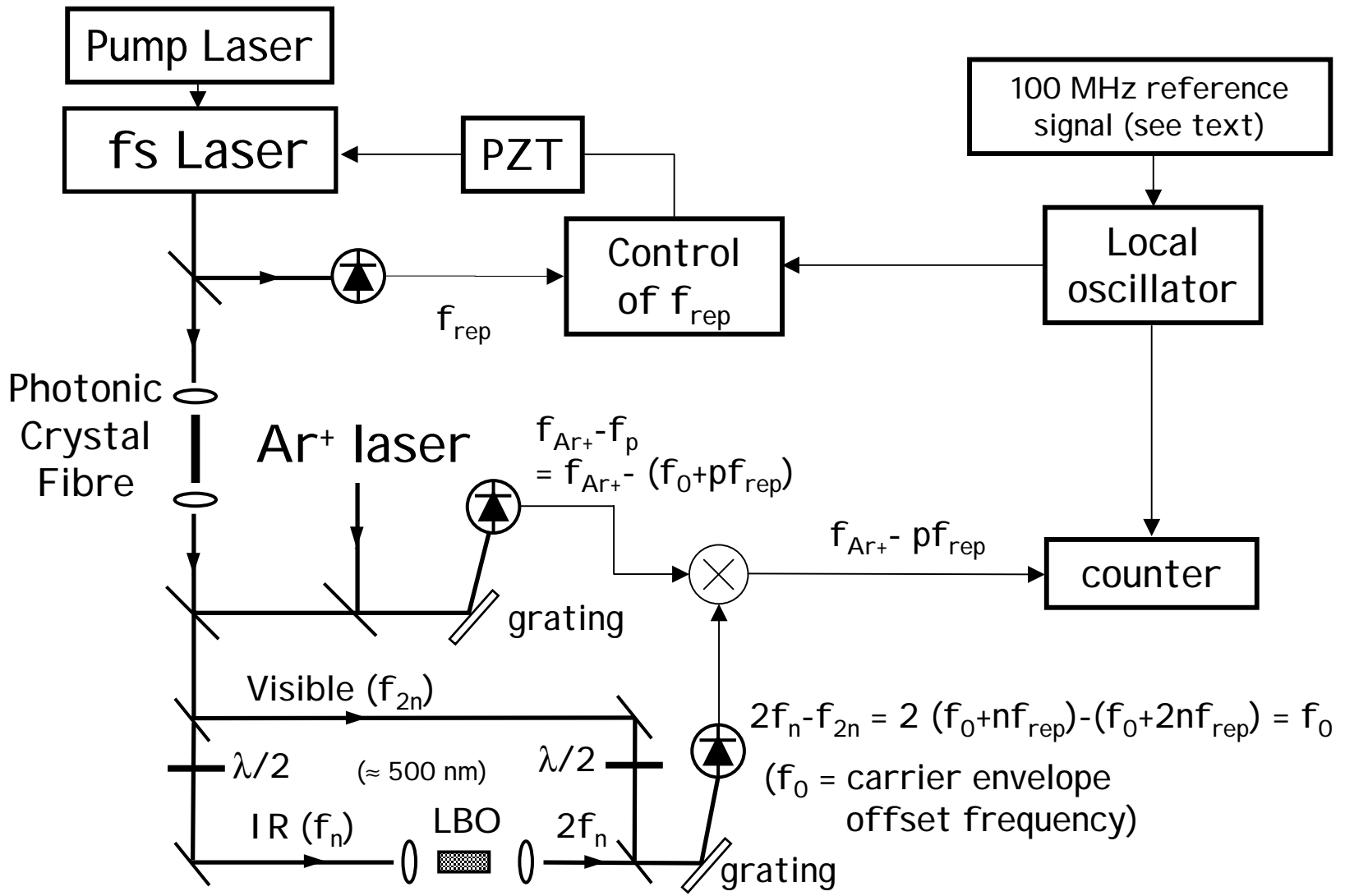

**Fig. 4**

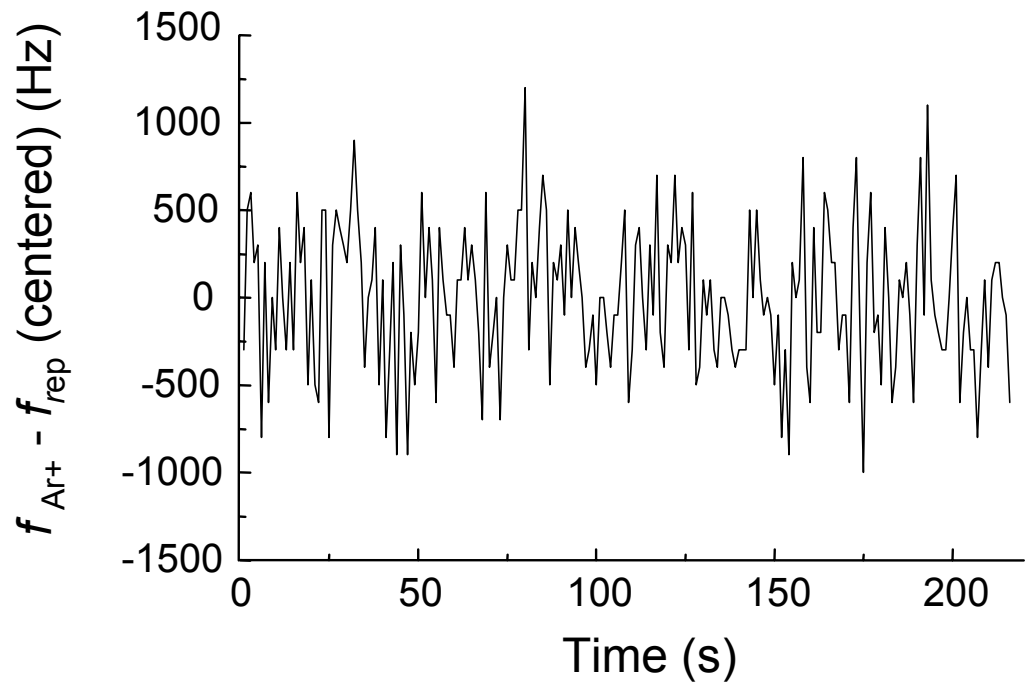

**Fig. 5**

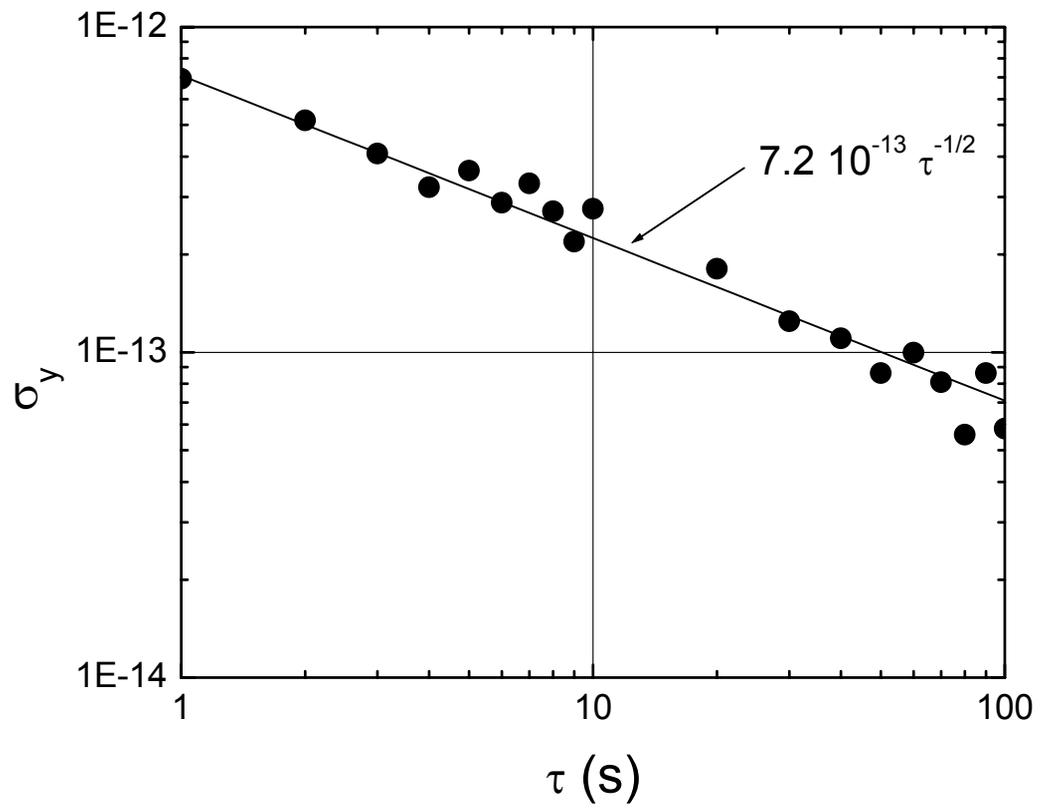

**Fig. 6**

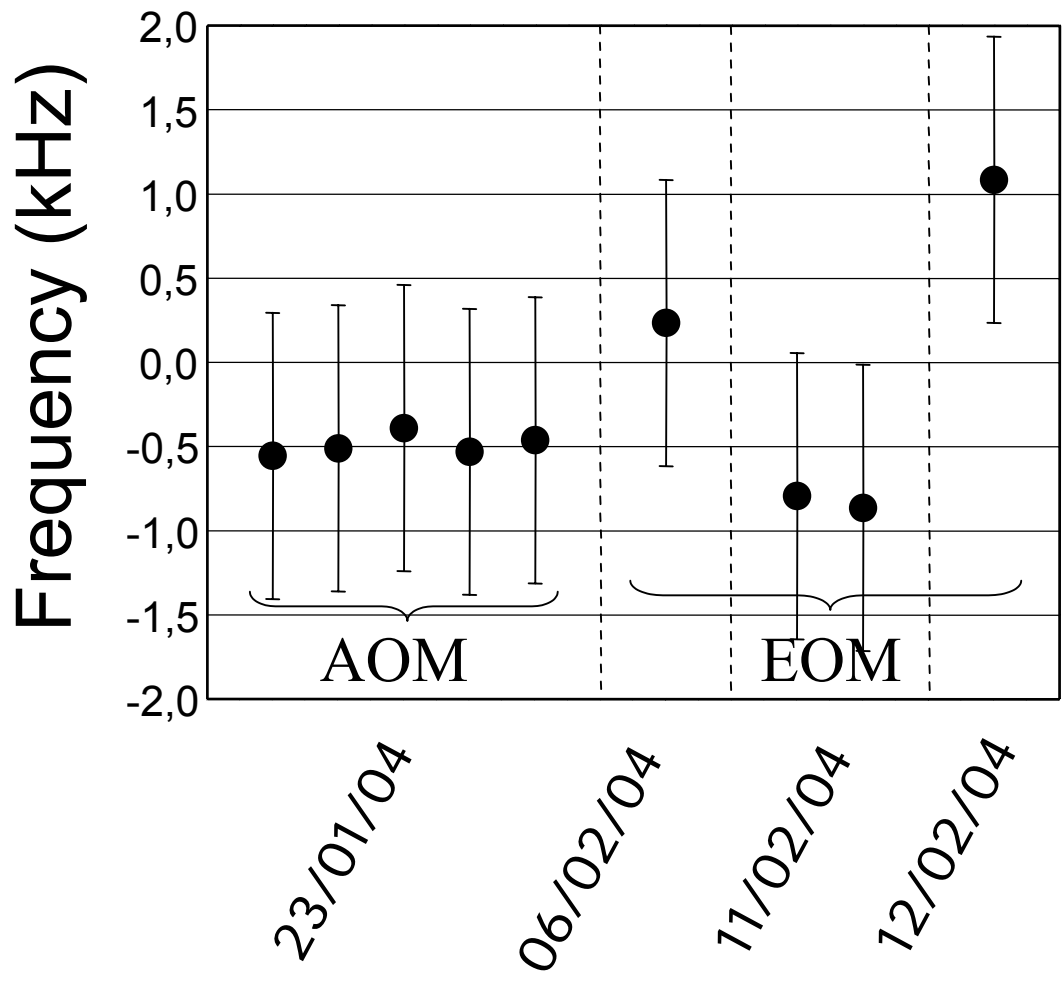

**Fig. 7**